\pdfoutput=1

\documentclass[aps,prl,showpacs,reprint]{revtex4-1}
\usepackage{graphicx}
\usepackage{subfigure}
\usepackage{amsmath}
\usepackage{amssymb}
\usepackage{dsfont}

\usepackage{color} 

\newcommand{\as}[1]{\textcolor{blue}{#1}}

\usepackage{hyperref}
\hypersetup{
colorlinks=true,final=true,
        linkcolor=blue,
        citecolor=blue,
        filecolor=blue,
        urlcolor=blue,
        pdfauthor={L. Steffen},
        pdftitle={Realization of Quantum Teleportation with Solid State Qubits}
}

\begin{document}

\title{Strong Coupling Cavity QED with Gate-Defined Double Quantum Dots\\
Enabled by a High Impedance Resonator}
\author{A.~Stockklauser}
\thanks{These authors contributed equally to this work.}
\author{P.~Scarlino}
\thanks{These authors contributed equally to this work.}
\author{J.~V.~Koski}
\author{S.~Gasparinetti}
\author{C.~K.~Andersen}
\author{C.~Reichl}
\author{W.~Wegscheider}
\author{T.~Ihn}
\author{K.~Ensslin}
\author{A.~Wallraff}
\affiliation{Department of Physics, ETH Zurich, CH-8093 Zurich, Switzerland}
\date{\today}
\begin{abstract}
The strong coupling limit of cavity quantum electrodynamics (QED) implies the capability of a matter-like quantum system to coherently transform an individual excitation into a single photon within a resonant structure. This not only enables essential processes required for quantum information processing but also allows for fundamental studies of matter-light interaction. In this work we demonstrate strong coupling between the charge degree of freedom in a gate-defined GaAs double quantum dot (DQD) and a frequency-tunable high impedance resonator realized using an array of superconducting quantum interference devices (SQUIDs). In the resonant regime, we resolve the vacuum Rabi mode 
splitting of size $2g/2\pi=238$\,\rm{MHz} at a resonator linewidth $\kappa/2\pi=12$\,\rm{MHz} and a DQD charge qubit dephasing rate of $\gamma_2/2\pi=80$\,\rm{MHz} extracted independently from microwave spectroscopy in the dispersive regime. Our measurements indicate a viable path towards using circuit based cavity QED for quantum information processing in semiconductor nano-structures.
\end{abstract}
\maketitle


In the strong coupling limit, cavity QED realizes the coherent exchange of a single quantum of energy between a nonlinear quantum system with two or more energy levels, e.g.~a qubit, and a single mode of a high quality cavity capable of storing individual photons \cite{Haroche2006}. The distinguishing feature of strong coupling is a coherent coupling rate $g$, determined by the product of the dipole moment of the multi-level system and the vacuum field of the cavity, which exceeds both the cavity mode linewidth $\kappa$, determining the photon life time, and the qubit linewidth $\gamma_2 = \gamma_1/2+\gamma_\varphi$, set by its energy relaxation and pure dephasing rates, $\gamma_1$ and $\gamma_\varphi$, respectively.

The strong coupling limit of Cavity QED has been reached with a multitude of physical systems including alkali atoms \cite{Thompson1992}, Rydberg atoms \cite{Brune1996}, superconducting circuits \cite{Wallraff2004,Chiorescu2004} and optical transitions in semiconductor quantum dots \cite{Yoshie2004a,Reithmaier2004}. Of particular interest is the use of this concept in quantum information processing with supercondcuting circuits where it is known as circuit QED \cite{Wallraff2004,Blais2004,Schoelkopf2008}.

Motivated by the ability to suppress the spontaneous emission of qubits beyond the free space limit \cite{Houck2008}, to perform quantum non-demolition (QND) qubit read-out \cite{Schuster2005,Wallraff2005}, to couple distant qubits through microwave photons coherently \cite{Majer2007,Sillanpaa2007} and to convert quantum information stored in stationary qubits to photons \cite{Houck2007,Eichler2012b},
research towards reaching the strong coupling limit of cavity QED is pursued for the charge and spin degrees of freedom in semiconductor nano-structures \cite{Delbecq2011,Frey2012,Petersson2012a,Toida2013,Wallraff2013,Viennot2015}.
Recently, in parallel with the work discussed here, independent efforts to reach this goal have come to fruition  with gate defined DQDs in silicon \cite{Mi2016a} and carbon nanotubes \cite{Bruhat2016b}.


The essence of our approach to reach the strong coupling limit with individual electronic charges in GaAs DQDs is rooted in the enhancement of the electric component of the vacuum fluctuations $\propto \sqrt{Z_r}$  \cite{Devoret2007} by increasing the resonator impedance $Z_r$ beyond the typical $50 \,\rm{\Omega}$ of a standard coplanar waveguide. We have realized a frequency-tunable microwave resonator with impedance $Z_r = \sqrt{L_r/C_r}\sim 1.8 \, \rm{k\Omega}$ using the large inductance $L_r\sim 50 \,\rm{nH}$ of a SQUID array \cite{Castellanos2007,Masluk2012,Altimiras2013}
combined with a small stray capacitance $C_r\sim 15 \,\rm{fF}$. Its resonance frequency and thus also its impedance is tunable by applying a small magnetic field using a mm-sized coil mounted on the sample holder. The frequency-tunability of the resonator is particularly useful in this context as it allows for the systematic study of its interaction with semiconductor nano-structures without changing their electrical bias conditions.

\begin{figure*}[!t]
\includegraphics[width=\textwidth]{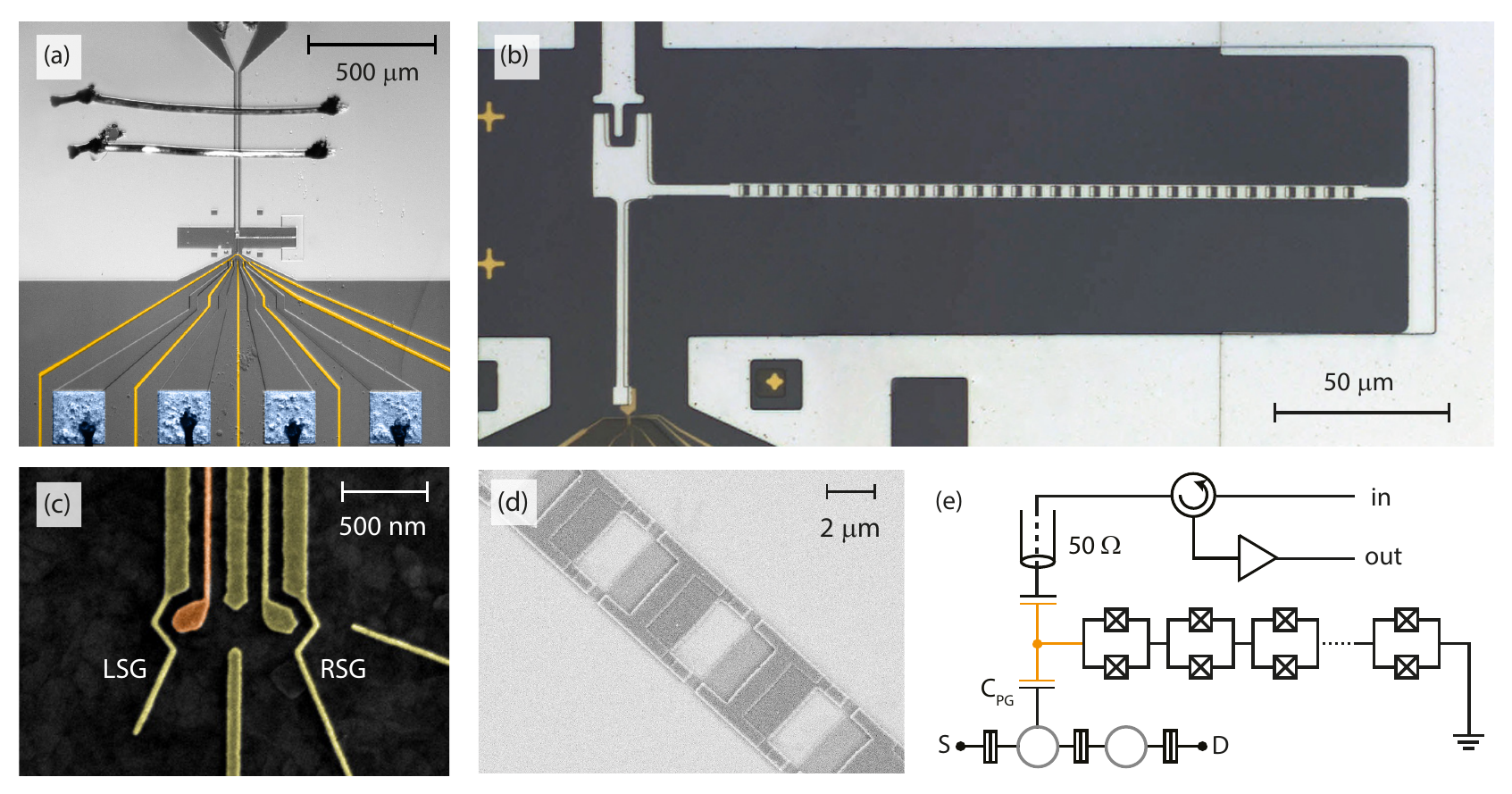}
\caption{Sample and simplified circuit diagram.
(a) False-color optical micrograph of a representative device indicating the substrate (dark gray), the superconducting structures (light gray), the gold top gates (yellow) forming the DQD and its source and drain leads and contacts (blue).
(b) Optical micrograph displaying a SQUID array resonator (light gray) and its coupling gate to the DQD and the DQD biasing structures (yellow).
(c) Electron micrograph of the DQD showing its electrostatic top gates (yellow) and the plunger gate coupled to the resonator (orange).
(d) Electron micrograph of three SQUID loops (dark grey) in the array deposited on the etched GaAs heterostructure (light gray).
(e) Circuit diagram schematically displaying the DQD  (source contact labeled S, drain contact labeled D, and coupling capacitance $C_\text{PG}$ to the resonator) and essential components in the microwave detection chain (circulator, amplifier) used for performing reflectance measurements of the device. Boxes with crosses and rectangles indicate Josephson and normal tunnel junctions, respectively.
}
\label{fig:SampleAndCircuit}
\end{figure*}

The resonator, with a small footprint of $300 \times 120 \,\rm{\mu m}^2$ (Fig.~\ref{fig:SampleAndCircuit}a,b), is fabricated using standard electron-beam lithography and shadow evaporation of aluminum (Al) onto a GaAs heterostructure. The embedded two-dimensional electron gas (2DEG) has been etched away everywhere but in a small mesa region hosting the DQD. The array, composed of $32$ SQUIDs (Fig.~\ref{fig:SampleAndCircuit}d), is grounded at one end and terminated in a small island at the other end to which a single coplanar drive line is capacitively coupled. A gate line extends from the island and forms one of the plunger gates of the double quantum dot (orange) (Fig.~\ref{fig:SampleAndCircuit}c).

The double quantum dot is formed in the mesa structure using gold (Au) top gates (yellow in Fig.~\ref{fig:SampleAndCircuit}a,b,c) controlling the tunnel coupling of the DQD to the source and drain leads (blue) as well as the inter-dot tunnel coupling $t$.
The left and right side gates (LSG, RSG) control the on-site electrostatic energies of each of the two dots, while the plunger gates are not biased in the experiment.
An additional gate and pair of leads can be configured as a quantum point contact for charge detection.
The microwave response of the system is probed in reflection (Fig.~\ref{fig:SampleAndCircuit}e) using standard circuit QED heterodyne detection techniques \cite{Frey2012,Wallraff2004}.

\begin{figure}[b]
\includegraphics[width=\columnwidth]{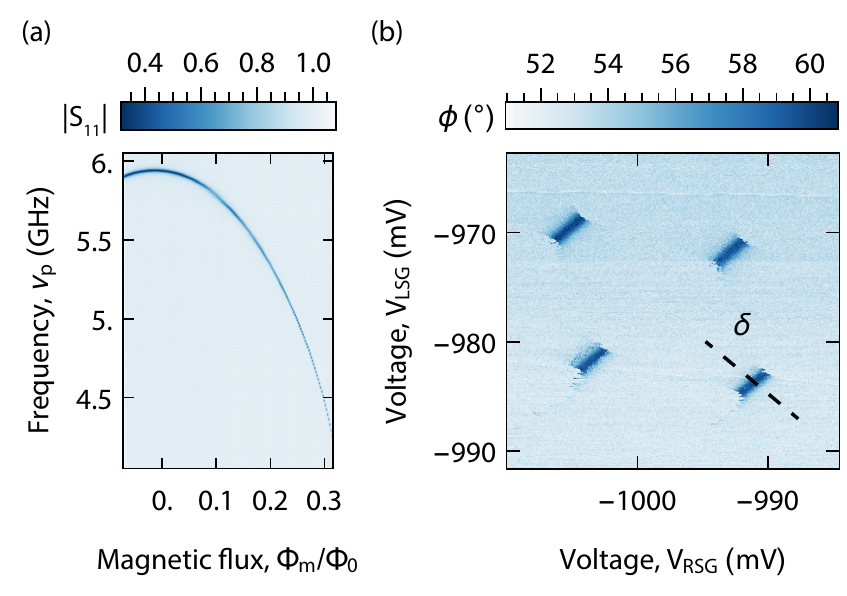}
\caption{Characterization of the SQUID array resonator and double quantum dot.
(a) Reflectance spectrum $|S_{11}|$ of the resonator as a function of probe frequency $\nu_p$ and applied magnetic flux $\Phi_\text m/\Phi_0$.
(b) Hexagonal charge stability diagram of the DQD detected in the phase $\phi$ of the microwave tone at frequency $\nu_p$ reflected of the resonator close to its resonance frequency $\nu_r$ as a function of the applied side gate voltages $V_\text{RSG, LSG}$.
}
\label{fig:ResDQDChar}
\end{figure}

We show that the resonance frequency of the SQUID array resonator can be tuned from a maximum value of $\nu_r\sim6.0\,\rm{GHz}$ to well below $4.5 \, \rm{GHz}$ (which is the lower cut-off frequency of our detection electronics) in measurements of its reflectance $|S_{11}(\nu_p)|$ as a function of applied magnetic flux $\Phi_\text m$ and probe frequency $\nu_p$ (Fig.~\ref{fig:ResDQDChar}a). From this data we extract the characteristic circuit parameters of the resonator and find that its impedance changes from $Z_r\sim 1.3\,\rm{k\Omega}$ to $1.8\,\rm{k\Omega}$ in this frequency range. With the DQD well detuned from the resonator biased at $\nu_r=5.02\,\rm{GHz}$, we determine its internal loss rate, its external coupling rate to the input line and the total line width
$(\kappa_{\rm{int}},\kappa_{\rm{ext}},\kappa)/(2\pi) \sim (10.0, 2.3, 12.3)\,\rm{MHz}$ \cite{Goppl2008}.

We configure the double quantum dot and determine its characteristic properties by extracting the amplitude and phase change of a coherent tone reflected off the resonator at frequency $\nu_p$ using a measurement of the reflection coefficient $S_{11}(\nu_p)$ in response to changes of the  potentials applied to the gate electrodes forming the double quantum dot. Using this by now well-established technique \cite{Delbecq2011,Frey2012,Petersson2012a}, we record characteristic hexagonal charge stability diagrams (Fig.~\ref{fig:ResDQDChar}b) from which we extract the DQD charging energy of $580\,\rm{GHz}$ and estimate the number of charges in each dot to be of the order of $10$ electrons \cite{Frey2012,Basset2013}.

\begin{figure*}[t]
\includegraphics[width=\textwidth]{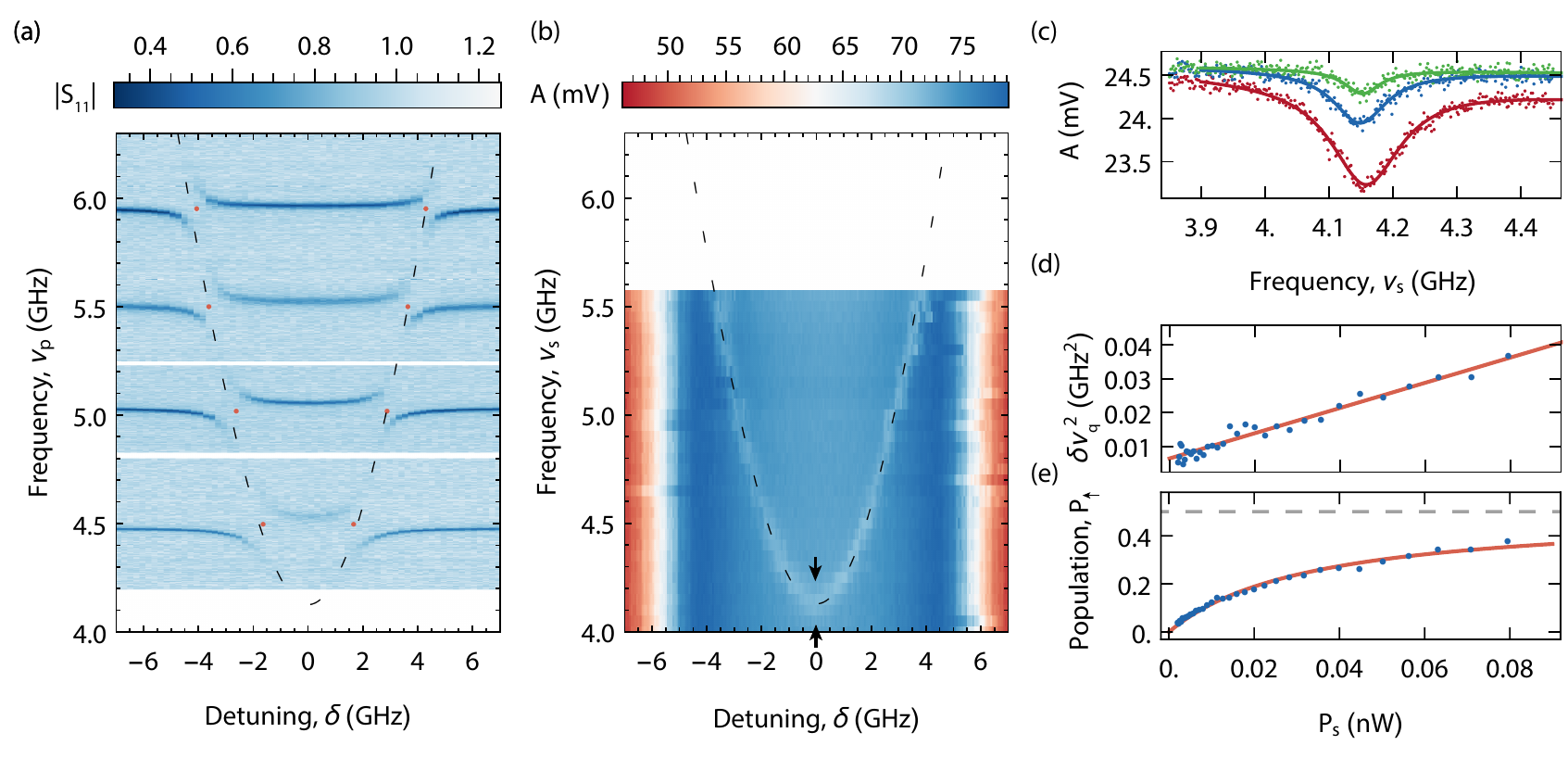}
\caption{DQD charge qubit spectroscopy.
(a) Resonator reflectance spectra $|S_{11}|$ as a function of probe frequency $\nu_p$ and DQD detuning $\delta$ for $\nu_r(\Phi_\text m) \approx \{4.5, 5.0, 5.5, 5.9\}\,\rm{GHz}$. Red points indicate resonance ($\nu_q = \nu_r$) extracted form the data. The dashed line indicates the calculated transition frequency of the charge qubit.
(b) Amplitude A of fixed frequency measurement tone $\nu_p=5.947\,\rm{GHz}$ reflected from the resonator \textsl{vs.}~qubit spectroscopy frequency $\nu_s$ and qubit detuning $\delta$. The dashed line indicates the expected qubit resonance frequency for $2t = 4.13 \, \rm{GHz}$.
(c) Qubit line shapes \as{$A(\nu_s)$} (dots) measured at $\delta=0$ (arrows in b) for for drive strengths $P_s = \{-5,-10,-15\}\,\rm{dBm}$ at the generator and fits to Lorentzian lines (on a linear background) extracting the linewidth $\delta\nu_q$. Probe frequency $\nu_\text p=5.022\,\rm{GHz}$, probe power $P_\text p=-35\,\rm{dBm}$ at the generator.
(d) Extracted qubit linewidth $\delta\nu_q^2$ (blue dots) \textsl{vs.}~spectroscopy drive power $P_s$ with linear fit (red solid line).
(e) Saturation of qubit population with spectroscopy drive power $P_s$.
}
\label{fig:QubitSpec}
\end{figure*}

To explore their mutual coupling, we first fix the SQUID array resonance frequency to $\nu_r = 5.03 \,\rm{GHz}$ and set the tunnel coupling of the DQD to $2t \sim 4.13\, \rm{GHz} < \nu_r$. This ensures that tuning the difference energy $\delta$ between the charge states in the right and left  quantum dot results in a resonance ($\nu_q = \nu_r$) between the charge qubit transition frequency $\nu_q$ and the resonator at $\delta_\pm = \pm \sqrt{(\nu_r(\Phi_\text m))^2-(2t)^2}$ \footnote{The conversion factor between the gate voltages and energy detuning $\delta$ has been estimated from the DQD spectroscopy measurements reported in Fig.~\ref{fig:QubitSpec}.}.

\begin{figure}[b]
\includegraphics[width=\columnwidth]{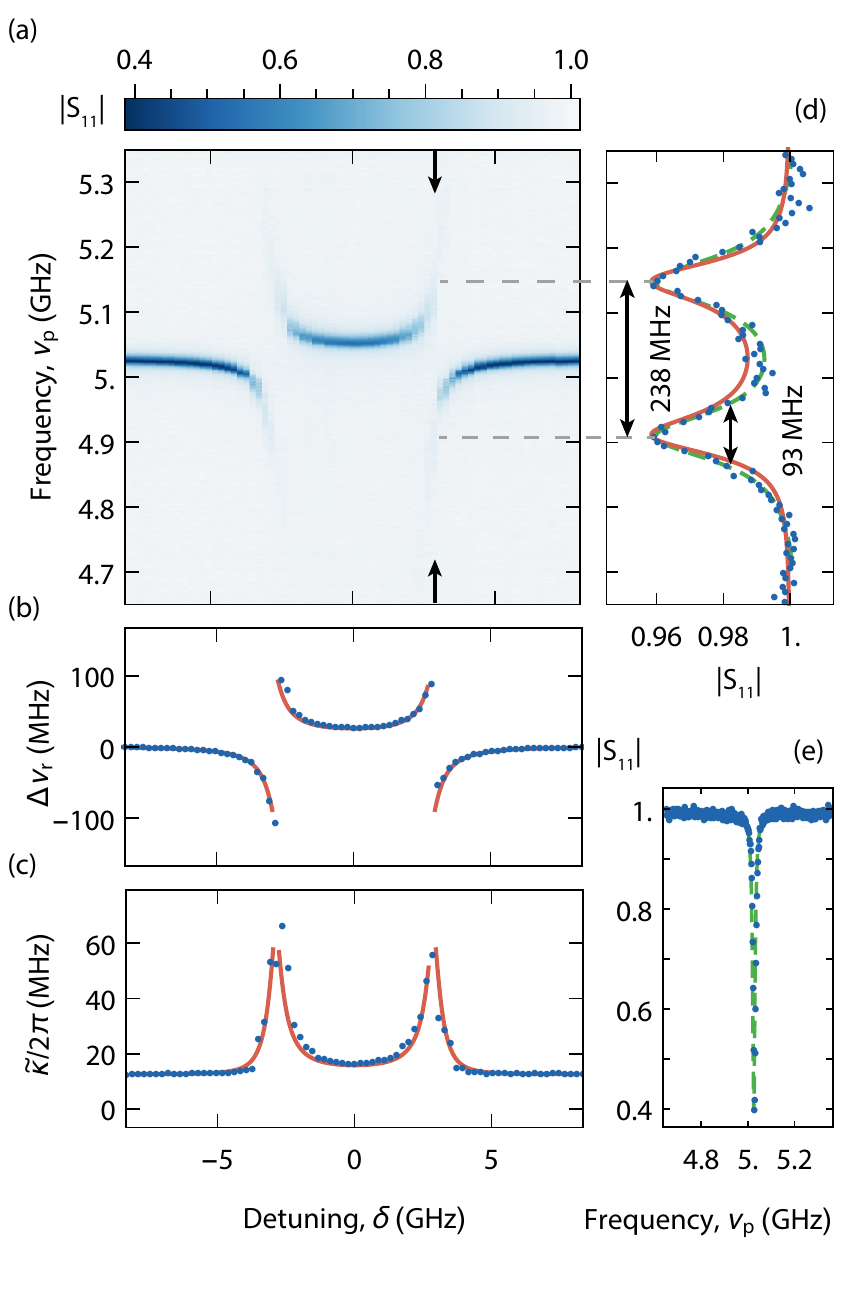}
\caption{Dispersive and strong resonant interaction.
(a) Resonator reflectance $|S_{11}|$ as a function of probe frequency $\nu_p$ and DQD detuning $\delta$. Resonance ($\nu_q=\nu_r$) occurring at $\delta_\pm$ is indicated by arrows.
(b) Extracted resonator frequency shift $\Delta \nu_r$ (dots) and
(c) linewidth $\tilde{\kappa}$ (dots) \textsl{vs.}~DQD detuning $\delta$ in comparison to results of a master equation simulation (line) for
$(g_0,\gamma_1^b,\gamma_\varphi^b)/(2\pi)= (155, 35, 63)\,\rm{MHz}$.
(d) Measured resonator reflectance $|S_{11}|$ (dots) \textsl{vs.}~probe frequency $\nu_p$ at resonance ($\nu_q=\nu_r$) displaying a strong coupling vacuum Rabi mode splitting. The solid line is the result of the master equation simulation, the dashed line is a fit to a superposition of two Lorentzian lines. (e) Resonator reflectance spectrum $|S_{11}|$ with a Lorentzian fit (dashed line) in the dispersive regime \textsl{vs.}~probe frequency $\nu_p$.
}
\label{fig:CavityQED}
\end{figure}

Varying the detuning $\delta$ (along the dashed line indicated in Fig.~\ref{fig:ResDQDChar}b) by applying appropriately chosen voltages to the two side gates we observe the dispersive (i.e.~non-resonant) interaction between the DQD and the resonator in a probe-frequency-dependent reflectance measurement of the resonator (Fig.~\ref{fig:CavityQED}a). As a function of $\delta$, the reflectance spectrum $|S_{11}(\nu_p)|$ shows characteristic shifts in the dispersive regime ($\nu_q \gg \nu_r$ or $\nu_q \ll \nu_r$) and indications of an avoided crossing at $\delta_\pm \sim \pm 2.86 \, \rm{GHz}$ at resonance ($\nu_q = \nu_r$) which we analyze in more detail below.

We first extract the frequency $\tilde{\nu}_r$ of the resonator, as renormalized by its interaction with the DQD, by fitting a Lorentzian line to the reflectance spectrum at each value of $\delta$. 
When varying $\delta$, the experimentally extracted shift $\Delta \nu_r = \tilde{\nu}_r - \nu_r$ reaches up to $\sim 100 \, \rm{MHz}$ close to resonance (blue dots, Fig.~\ref{fig:CavityQED}b). The measured values of $\Delta \nu_r$ are in excellent agreement with the results of a master equation simulation (solid line) analyzed in the same way finding the parameters
$(g_0,\gamma_1^b,\gamma_\varphi^b)/(2\pi)= (155, 35, 63)\,\rm{MHz}$
while keeping the bare resonator linewidth $\kappa$ fixed at its independently determined value stated above.
In the Jaynes-Cummings model we use to describe the coupled system, both the coupling rate and the decoherence rates depend on the mixing angle $\theta$. The effective coupling strength $g$ is given by $g=g_0\sin\theta$, where $\sin\theta=2t/\sqrt{(2t)^2+\delta^2}$, while the decay and decoherence rates are given by
$\gamma_1 = \sin^2\theta \gamma_\varphi^b + \cos^2\theta \gamma_1^b$ and
$\gamma_\varphi = \cos^2\theta \gamma_\varphi^b + \sin^2\theta \gamma_1^b.$
%
Using the same set of parameters we also find excellent agreement with the effective linewidth $\tilde{\kappa}$ of the resonator as renormalized by the hybridization with the DQD charge qubit. Detuned from the quantum dot, the resonator displays the bare linewidth $\kappa$. When approaching resonance, it is increased by more than a factor of $4$ due to the interaction with the qubit with significantly larger linewidth $\gamma_2\gg\kappa$. Near resonance $\nu_q \sim \nu_r$
the resonator reflectance does not display a single Lorentzian line shape in probe frequency but develops two well resolved spectral lines.

Tuning the DQD into resonance with the resonator ($\nu_q = \nu_r$), indicated by arrows in Fig.~\ref{fig:CavityQED}a, we observe a clear vacuum Rabi mode splitting (blue dots) in the reflectance spectrum of the resonator (Fig.~\ref{fig:CavityQED}d). A fit (dashed green line) of the spectrum to a superposition of two Lorentzian lines yields a splitting of $2g/2\pi\sim238\,\rm{MHz}$, with an effective linewidth of $93\,\rm{MHz}$.
The vacuum Rabi mode splitting is found to be in good agreement with the spectrum evaluated from the master equation simulation (red solid line) with the parameters
$(g_0,\gamma_1^b,\gamma_\varphi^b)/(2\pi)= (155, 35, 63)\,\rm{MHz}$,
which is consistent with the analysis of the dispersive frequency shift discussed above.
We note that the small amplitude of the signal in reflection is a direct consequence of the fact that the qubit decoherence rate $\gamma_2$ is significantly larger than the resonator decay rate $\kappa$, an observation that is also reproduced in the theoretical analysis of the data.


Furthermore, we analyze the spectroscopic properties of the DQD charge qubit in two complementary measurements. First, we make use of the frequency tunability of the high impedance SQUID array resonator by applying a small magnetic flux $\Phi_\text m$ to its SQUID loops and keeping the DQD charge qubit at a fixed tunnel coupling $2 t$. At a set of frequencies $\{\nu_r(\Phi_m)\}$, we observe resonator spectra characteristic for its dispersive and resonant interaction with the qubit (Fig.~\ref{fig:QubitSpec}a). The resonances ($\nu_q = \nu_r$) occurring at $\delta_\pm$ for the set of values $\{\nu_r(\Phi_\text m)\}$ (red data points) are in good agreement with the expected dependence of the qubit energy levels on $\delta$, see dashed line in Fig.~\ref{fig:QubitSpec}a. We note that at each resonance ($\nu_q = \nu_r(\Phi_m)$) an avoided crossing displaying a vacuum Rabi mode splitting is observed.

We also perform qubit spectroscopy by probing the amplitude and phase of the resonator reflectance at fixed measurement frequency $\nu_p = 5.947 \, \rm{GHz}$ while
applying an additional spectroscopy microwave tone at frequency $\nu_s$ to the resonator. When the spectroscopy tone is resonant with the qubit transition frequency ($\nu_s = \nu_q$) the qubit is excited from its ground state $|g\rangle$ to a mixture between ground and excited state $|e\rangle$. This mixed state changes the resonance frequency $\tilde{\nu}_r$ of the resonator by dispersive coupling resulting in a detectable change of the amplitude A (and also phase $\phi_r$, not shown) of the microwave tone reflected at frequency $\nu_p$ (Fig.~\ref{fig:QubitSpec}b).
This technique has been pioneered for superconducting qubits \cite{Schuster2005,Wallraff2005} where it is widely used. Varying both the qubit detuning $\delta$ and the spectroscopy frequency $\nu_s$ we map out the spectrum of the qubit (dashed line, Fig.~\ref{fig:QubitSpec}b) and determine its tunnel coupling $2 t = 4.13 \, \rm{GHz}$ .

Using this technique we are not only able to accurately determine the transition frequency $\nu_q$ of the DQD charge qubit but also its line shape, shown for three drive powers $P_s$ in Fig.~\ref{fig:QubitSpec}c. The observed line shape depends on the qubit intrinsic linewidth, as set by its dephasing time $T_2^{\star}$, and on the strength of the applied microwave drive $P_s$ which broadens the line proportional to its amplitude. In the limit of weak driving ($P_s \rightarrow 0 $), the spectroscopic linewidth $\delta \nu_q \sim 80 \, \rm{MHz}$ is determined by the dephasing time $T_2^{\star} = 12.5 \, \rm{ns}$ of the DQD qubit as extracted from a linear fit to the data in Fig.~\ref{fig:QubitSpec}d. This is consistent with the previously extracted values of $\gamma_2$. Increasing the drive strength $P_s$ we observe the qubit transition and thus also the resonator response to approach saturation (Fig.~\ref{fig:QubitSpec}e).


The data presented in this manuscript indicates that the strong coupling limit of a semiconductor charge qubit formed in a double quantum dot coupled to a microwave photon has been realized. This is achieved by the use of a high impedance SQUID array resonator increasing the coupling strength by a factor of $6$ relative to coupling schemes using conventional $50 \, \rm{\Omega}$ resonators. This approach is universally applicable to any circuit QED application striving to maximize the coupling to the charge degree of freedom.
The realization of strong coupling in this semiconductor circuit QED device also enabled us to perform spectroscopy of the DQD qubit in the dispersive regime to evaluate its line shape in dependence on the microwave drive power, indicating the possibility of temporally resolving the charge dynamics. These results carry promise to further advance quantum information processing efforts based on semiconductor charge and spin qubits using circuit QED approaches, e.g.~to perform quantum non-demolition (QND) readout and to realize coupling between distant qubits through microwave photons.


We acknowledge contributions by Michele Collodo, Andreas Landig, Ville Maisi and Anton Poto\v{c}nik. We thank Alexandre Blais for valuable feedback on the manuscript. 
This work was supported by the Swiss National Science Foundation through the National Center of Competence in Research (NCCR) Quantum Science and Technology and by ETH Zurich.

\bibliographystyle{apsrev4-1}

%

\end{document}